\documentclass[pre, twocolumn, amsmath,superscriptaddress]{revtex4}
\usepackage{dcolumn}
\usepackage{bm}
\usepackage{graphicx}
\usepackage{color}
\usepackage{xcolor}
\usepackage{amsmath}
\usepackage{amssymb}
\usepackage{dcolumn}
\usepackage{epsfig}
\usepackage{hyperref}

\DeclareGraphicsExtensions{. jpg,. pdf, . mps, . png, . eps, . ps, . EPS}

\newcommand{\mediaT}[1]{\left\langle #1 \right\rangle}

\def\defi{{\buildrel \;def\; \over =}}
\newcommand{\be}{\begin{equation}}
\newcommand{\ee}{\end{equation}}

\newcommand{\media}[1]{\langle #1 \rangle}

\def\mod{ \mathop{\rm mod} }

\definecolor{myred}{RGB}{168,5,14}
\definecolor{myblue}{RGB}{13,13,255}
\definecolor{editorcolor}{RGB}{200,5,14}
\definecolor{mygreen}{RGB}{20,150,20}

\begin{document}

\title{Partitioning networks into clusters of synchronized nodes via 
the message-passing algorithm: 
an unbiased scalable {approach}
}

\author{Massimo Ostilli}
\affiliation{
Instituto de
    F\'isica, Universidade Federal da Bahia, Salvador, {BA 40170-115,} Brazil}

\begin{abstract}
  Partitioning large networks into stable clusters of synchronized nodes is a challenging task.
  Recent approaches based on spectral analysis can provide exact results on specific dynamics but remain unfeasible
  for very large networks.
  Moreover, within a stochastic framework, it is unclear which dynamics should be chosen to study synchronization.  
  Here we propose an unbiased and scalable method based on the message-passing algorithm.
  By exploiting the collective behavior emerging across critical points of an effective Ising-like model, we identify dynamically coherent clusters of synchronized nodes
  and illustrate the approach on some large real-world networks.
  We find that, unlike continuous-time dynamics, abrupt desyncrhronization occurs even in simple graphs,
  without the need to invoke higher order interactions. However, when noise
  is included, the transition to synchronization becomes smoother and proceeds through the formation of plateaus,
  albeit at the cost of requiring larger coupling strengths.
\end{abstract}


\maketitle


\section{Introduction}
Network theory~\cite{Barabasi2016,Newman2018,DorogovtsevMendes2022} provides a universal framework for investigating complex systems across a wide range of domains, including physical, biological, ecological, and social systems, and at different levels of organization, from structural to functional. Over the past three decades, driven by the digital, internet, and more recently AI revolutions, the theory of complex networks has experienced a rapid expansion in both theoretical developments and applications. This growth has been fueled by the availability of massive real-world datasets, whose size and heterogeneity have themselves become a major challenge, commonly referred to as “Big Data”~\cite{dean2008,mckinsey2011,gandomi2015}.

In many scientific fields, the volume and diversity of available data now exceed the capabilities of traditional numerical methods.
This issue is particularly severe in biological and ecological systems~\cite{kelling2009,reichman2011,marx2013,stephens2015},
where data are produced across multiple spatial and temporal
scales and by heterogeneous experimental and observational techniques. Genomics is a paradigmatic example, with genome sizes ranging from 
$10^4$ base pairs to several billions in humans. In such contexts, a direct or naive application of network theory rapidly becomes impractical.

One of the original motivations for adopting a network-based description lies in its ability to reduce the full
$N\times N$ connectivity information of a system to a compact statistical characterization in terms of a limited set of global metrics, such as the degree distribution, clustering coefficient, and average path length. These quantities capture essential features of the system and play a central role in the analysis of dynamical processes on networks,
such as
percolation and epidemic spreading~\cite{Review}. 

Beyond global descriptors, local structural properties are also crucial. In this respect,
the older notion of community~\cite{Barabasi_Hierarchical,Santo,Santo2}---loosely defined as a subset of elements exhibiting similar behavior—has long played a central role.
It turns out that the vague idea of similarity among a group of nodes can be made precise by using
the more recent concept of symmetry of the labeled network,
for which several closely related definitions exist~\cite{MacArthur,Pecora,Lin,Lambiotte,Fiore,Gambuzza,Sanchez,OstilliTE}.
Following Ref.~\cite{OstilliTE}, we have a symmetry when a subset of vertices have identical external neighborhoods and equal degree.

A subset with such a symmetry forms a group of topologically equivalent (TE) nodes whose dynamics, under uniform initial conditions, evolve coherently. 
Depending on the nature of the dynamics, this coherence --- if stable --- may manifest as convergence to a common stationary state (consensus) or as sustained coherent oscillations (synchronization). In both cases, the underlying mechanism is the same.
Consequently, identifying partitions of a network into symmetric groups provides valuable insight for the network functionality.
Conversely, the observed collective dynamics can be exploited to infer structural properties of the underlying network.
The present work goes in the latter direction.

In recent years, the general theory of synchronization has undergone significant advances, with important applications to complex real-world networks. A key outcome of these developments is a twofold understanding: first, the above mentioned central role of network symmetries
in enabling cluster-~\cite{Pecora,Lin,Lambiotte,Fiore,Gambuzza,Multilayer} and protected-synchronization~\cite{OstilliTE};
second, the empirical observation that real-world networks, unlike traditional models of network theory
(e.g., Random-Graph, Configuration, or Barabási–Albert models), typically exhibit a large abundance of such symmetries~\cite{MacArthur,Sanchez}.
Despite these progresses, their practical implications seem only partially explored and rarely exploited in applied network analysis, as noted in Ref.~\cite{Sanchez}.

Two main reasons may account for this gap. On the one hand, existing approaches to study local synchronization rely heavily on spectral methods, which become computationally prohibitive for large networks. On the other hand, in stochastic settings typical of biological and ecological systems, the relevant underlying dynamics is often unknown or poorly defined, making dynamical-model–based analyses difficult to implement.

In this work, we address these limitations by introducing a synchronization-detection framework based
on the message-passing algorithm (MPA)~\cite{Yedidia,Yedidia2,Mezard_Parisi,Mezard_Montanari}. By reducing the state variables to binary (Ising-like) degrees of freedom,
the MPA provides a scalable and unbiased surrogate dynamics that allows to identify the clusters of synchronized nodes without assuming any specific microscopic dynamical model.
The scalability of the approach follows from the near-linear growth of the computational cost with the network size $N$ {of the MPA}.

By varying the coupling parameter of the effective Ising model, we probe collective behavior across critical points and use it to discriminate clusters of synchronized nodes. We show that TE nodes act as nucleation centers for synchronization, giving rise to rich synchronization scenarios in networks with many symmetry-induced structures. In particular, we find that abrupt desynchronization transitions can occur even in simple graphs, without invoking higher-order interactions~\cite{Arenas_Higher_Order}.
When stochastic noise is introduced, however,
the transition becomes smoother and proceeds through well-defined plateaus, albeit at the cost of larger coupling strengths.

The paper is organized as follows. Section II motivates the use of the MPA as an unbiased surrogate dynamics. Section III defines the observables used to characterize synchronization. Section IV formalizes the notion of topological equivalence. Section V reviews the message-passing algorithm. Section VI presents three variants of the proposed method, corresponding to uniform initial conditions, random initial conditions, and random initial conditions with noise. Section VII illustrates the approach on two large real-world networks. Conclusions and perspectives are given in Section VIII, followed by an appendix.

\section{\label{Unbiased}The message passing algorithm as an unbiased dynamic}
Most real-world complex systems are governed by stochastic dynamics arising from interactions among their elements, whose topology can be represented by a graph 
$\mathcal{G}=(V,E)$, which we hereafter assume to be static, as well as by interactions with the external environment. Even when the system is treated as autonomous—so that external influences are neglected or modeled as noise—the precise nature of these interactions and the resulting dynamics generally remains unknown. With the exception of purely physical or engineered systems, such as mechanical oscillators, electronic circuits, or power grids, the choice of a specific dynamical model is therefore often dictated by qualitative or phenomenological considerations.

Within the study of synchronization (or consensus), two main classes of models are commonly employed~\cite{StrogatzC,Pikovsky,ArenasReview}:
(i) Laplacian-like models and (ii) Kuramoto-like models. In both cases, the state of each node is updated based on the states of its neighbors
through either the graph Laplacian or the adjacency matrix of $\mathcal{G}$.
For example, assuming $N$ scalar state variables $(x_1,\ldots.x_N)$, representing the phases of $N$ ``oscillators'',
and a discrete-time dynamics occurring at times $t\in \mathbb{N}$,
in (i) we have $x_i(t+1)=x_i(t)+F_i(x_i(t))+ J \sum_j L_{i,j} G_j(x_j(t))$; and
in (ii) we have $x_i(t+1)=x_i(t)+\omega_i+ J \sum_j A_{i,j} S_{i,j}(x_j(t)-x_i(t)), ~\mod 2\pi$,
where $J$ is a global coupling; $L$ and $A$ are the Laplacian and adjacency-matrix of $\mathcal{G}$,
respectively; $F_i(\cdot)$, $G_i(\cdot)$, $S_{i,j}(\cdot)$ suitable functions; and $\omega_i$ natural frequencies.
In particular, when the focus is about global synchronization,
and the system consists in $N$ homogeneous oscillators (so that $F_i=F$, $G_i=G$, $S_{i,j}=S$, and $\omega_i=\omega$),
the Laplacian property $\sum_jL_{i,j}=0$ in (i) and the assumption $S(\cdot)=\sin(\cdot)$ in (ii) (justified by a perturbative analysis around the limit cycle~\cite{Pikovsky}),
guarantee the existence of the invariant synchronized solution $x_1(t)=\ldots=x_N(t)$ and non trivial questions concern its stability.
For (i) this stability analysis is provided by the ``master-stability function'', which has been consolidated as a standard method that becomes particularly interesting
when the state  variables $x_i$ are not scalars but $m$-dimensional vectors and $G(\cdot)$ couples non linearly the internal components of each vector $x_i$.
In (ii), instead, due to the intrinsic
non linear coupling among the oscillators, apart from the mean-field case, the stability analysis turns out to be much harder and a general receipt is still missing.
In both cases, however, the stability of the synchronous solution strongly depends on the spectral properties of the graph ($L$ for (i), or
$A$ for (ii)), but also on the other above model-specific assumptions (about $J$, $F$, $G$ for (i), or about $\omega$, $J$, $S$, for (ii)). 
The same considerations apply to the more interesting and general case of heterogeneous oscillators, each having its own natural frequency $\omega_i$. In this case, 
a much richer scenario with local synchronizations can set in. Here, by local synchronizations we mean that these synchronizations take place
within small clusters of similar oscillators, \textit{i.e.}, oscillators having close natural
frequencies but also similar topologies. As we shall see, this topic is central for the present work.  

However, the problem is not limited to the choice of a specific dynamical rule. A more fundamental issue concerns internal consistency.
If the actual dynamics governing the system is unknown, or if it is affected by stochastic fluctuations,
then the system should be treated probabilistically. At any given time $t$, the $N$ state variable $x_1(t),\ldots,x_N(t)$
should therefore be regarded as realizations of random variables.

\textit{Hereafter, we adopt the convention that $x_i$ denotes the random variable associated with node $i$, whereas $x_i(t)$ 
denotes its realization at time $t$.}

Let $P^{t}(x_1,\ldots,x_N)$ denote the joint probability distribution of the $N$ state variables at time $t$.
Regardless of how complicated the underlying microscopic dynamics may be, this distribution exists and can, in principle,
be estimated empirically from repeated realizations of the system with different initial conditions and, possibly, different noise realizations.
Assuming Markovianity, the time evolution of $P^{t}(x_1,\ldots,x_N)$
could be formally described by a master equation with suitable transition probabilities, which, however, we suppose to be unknown.
Importantly, we do not wish to introduce ad \textit{hoc} assumptions of the type underlying Laplacian-like or Kuramoto-like models.

Consider now a node $i$ and its set of neighbors $\mathcal{N}(i)$. Let us define $P^{t}_{i,\mathcal{N}(i)}\left(x_i|\{x\}_j,j\in\mathcal{N}(i)\right)$,
which describes the probability of $x_i$
conditioned on the states of its neighbors at the same time $t$, in formula
\begin{eqnarray}
  && P^{t}_{i,\mathcal{N}(i)}\left(x_i|\{x\}_j,j\in\mathcal{N}(i)\right)=\nonumber \\
  && \frac{\sum_{\{x_k\}:~k\neq i; k\notin \mathcal{N}(i)} P^{t}(x_1,\ldots,x_N)}
  {\sum_{\{x_k\}:~k\neq i} P^{t}(x_1,\ldots,x_N)}.
\end{eqnarray}  
We emphasize that this conditional probability does not coincide with the actual transition probability of the dynamics,
since, here, all variables are evaluated at the same time $t$. Nevertheless, these conditional probabilities can be used to construct an
artificial, surrogate dynamics whose purpose is not to reproduce the microscopic evolution of the system, but to probe its synchronization properties.

In fact, by means of a proper iterative procedure, the collection of the $P^{t}_{i,\mathcal{N}(i)}$'s can be used to reconstruct the 
the full joint distribution $P^{t}(x_1,\ldots,x_N)$. This allows one to investigate synchronization
by identifying stable clusters of nodes whose state variables evolve coherently along the iterative dynamics reproducing 
$P^{t}(x_1,\ldots,x_N)$, independently of the initial conditions.
Within certain limits, discussed later, this surrogate dynamics is naturally implemented by the message-passing algorithm (MPA),
also known as belief propagation. As the name suggests, the MPA operates through the propagation of suitable messages along the edges of the network.

Moreover, if we make the assumption (or reduction) to consider the random variables as Ising variables, $x_i\in\{-1,1\}$, as we shall see via Eqs. (\ref{pairs})-(\ref{pairs1}),
all the knowledge of the $P^{t}_{i,\mathcal{N}(i)}$'s can be entirely encoded via $|E|$ coupling values $J_{i,j}$ associated to the $|E|$ links of the given graph $\mathcal{G}$.
The use of binary state variables should be regarded as a coarse-grained description, whose implications for the resolution of synchronized clusters are discussed in the Conclusions.
Regardless of that, in this formulation, no specific microscopic dynamics is assumed: the only unknown quantities are the effective couplings $J_{i,j}$,
while no deterministic or stochastic update rule for the state variables is postulated. From this perspective, the procedure is unbiased with respect to the choice of dynamics.

Once the distribution of the messages has been achieved, the mean value of the state variables (or any function of them) can then be easily computed.
We shall review the MPA in detail in the next sections, but here we want to stress the following\\
\textit{\textbf{Key principle}:
  The dynamics that reproduces the distribution of the state variables starting from the conditional probabilities must be formulated as a local update of the messages, not as a local update of the state variables. Only in the limit of vanishing fluctuations—such as in fully connected graphs (mean-field limit) or in deterministic settings—do the
  two procedures become equivalent.}

This principle clarifies why, when the actual dynamics is unknown, assuming a local update of the state variables may introduce bias and, in many cases, lead to an overestimation of phase transitions, as is well known from the theory of critical phenomena.

\section{\label{Target}The target: partitioning into clusters of synchronized nodes}
Given a simple undirected graph $\mathcal{G}=(V,E)$ having $N=|V|$ labeled nodes with labels in the natural set  $\{1,\ldots,N\}$,
characterized by the symmetric adjacency-matrix ${A}$, and given a threshold $\epsilon>0$,
our target is to find, for any value of the coupling $J$, a partition of $V$ into $Q$ disjoint non empty sets (or clusters)
$H^{(1)},\ldots,H^{({Q})}\subset V$ such that, in each cluster all the nodes are synchronized up to the threshold $\epsilon$.
In other words, there exists ${Q}$ sets of indices $H^{(1)},\ldots,H^{({Q})}\subset V$ such that 
\begin{eqnarray}
  \label{TopoC}
    \cup_{l=1}^{Q} H^{(l)}=V, \quad H^{(l)}\cap H^{(m)} =\emptyset \quad \forall l\neq m,
\end{eqnarray}
and, for each $l$,
\begin{eqnarray}
  \label{dist}
  |x_i-x_j|<\epsilon, \quad \forall i,j\in H^{(l)}.
\end{eqnarray}
Note that, in Eq. (\ref{dist}), the case $i=j$ is allowed. This implies
that some of the sets $H^{(l)}$ can contain just one element.
By construction, such a partition always exists and is unique.

We anticipate that the above $\epsilon$ will be set to the smallest possible value compatibly
with the available machine precision of the specific computer where our algorithm will be run.
We shall not investigate synchronization versus threshold: a finite $\epsilon$ is merely
used to implement the algorithm numerically.
We shall investigate the above target with and without noise.

\section{Topologically equivalent nodes}
In parallel with the notion of clusters of synchronized nodes, there is the previously mentioned notion
of groups of TE nodes~\cite{OstilliTE}. Consider a set of $N'$ nodes with equal degree and, for each node, consider the sets of first neighbors that do not belong to the group. 
If these sets are all equal to each other, we say that the $N'$ nodes form a TE group.
In formulas, for $\forall i_1,\ldots,i_{|H|}\in H$, we want:
\begin{eqnarray}
  \label{Topo1}
  &&  \sum_{j\in V} {A}_{i_1,j}=\ldots = \sum_{j\in V} {A}_{i_{|H|},j}, \\
    \label{Topo2}
&& {A}_{i_1,k}=\ldots={A}_{i_{|H|},k}, ~ \forall k\in V\setminus H.
\end{eqnarray}
In other words, each node of the group sees both the same external labeled graph and the same internal non labeled graph
(\textit{i.e.}, internally, each node sees the same structure, but not the same nodes). Alternatively, one can say that $[A,P]=0$,
where $P$ is the permutation operator acting on the nodes of the given group.

Given a TE group, its nodes share the same graph topological features. 
The internal and external degree are just two examples of topological features shared by all the nodes of the group.
One should keep in mind that this holds for any other topological features. Think for example to the shortest path between any pair of nodes with one node
belonging to group $l$ and the other one belonging to group $m$ (possibly with $l=m$): it depends only on the group-indices $l$ and $m$, not on the specific
selected pair of nodes. Similarly, think to the number of paths of fixed length between any pair of nodes with one node
belonging to a group and the other one belonging to another group; etc...

Any graph owns its partition into, say, $Q_{\mathrm{TE}}$ groups of TE nodes and,
due to above features, groups of TE nodes represent good candidate as possible clusters of synchronized nodes.
As we shall see, however, this correspondence remains partial and limited to the case without noise.
Note that this observation holds in general for any symmetry (in the broader group-theoretical sense of ~\cite{Sanchez}):
symmetric clusters are good candidate as sets of synchronized nodes, however, whether they are stable or not
(under random initial conditions and without noise) depends on the entire spectra of the adjacency-matrix or of the Laplacian of $\mathcal{G}$~\cite{Boccaletti3}.

Yet, as we shall show by applying the MPA to synthetic graphs with a known partition in $Q_{\mathrm{TE}}$ groups of TE nodes,
it remains true that $Q$ and $Q_{\mathrm{TE}}$ have the same order of magnitude and, most importantly,
we shall see that TE groups act as nucleation centers for synchronization. 

\section{\label{MPAh} MPA}
  In this Section, we review the MPA (or belief-propagation algorithm)~\cite{Yedidia,Yedidia2,Mezard_Parisi,Mezard_Montanari}, which can be seen as
  an effective algorithm governing the propagation of messages through a physical or abstract network.
  More precisely, as anticipated in Sec. \ref{Unbiased}, the MPA reproduces all the features of the joint
  distribution $P^{t}(x_1.\ldots,x_N)$ from the knowledge of the conditional probabilities associated to the given graph $\mathcal{G}=(V,E)$. 
  The MPA finds application in countless diverse phenomena like physics, biology, and computer science. 
In the MPA, each link $(i,j)$ is associated with two directional messages, $u_{i\to j}$ and $u_{j\to i}$  (\textit{i.e.}, in general $u_{i\to j}\neq u_{j\to i}$ even though we have assumed $(i,j)$ to be undirected). The MPA local update of $u_{i\to j}$ ($u_{j\to i}$)
depends only from the messages arriving from the neighbors of $i$ ($j$), with $j$ excluded ($i$ excluded).
From now on, for the random variables we assume $x_i\in\{-1,1\}$ and we assume a partial independence from the conditioning variables as follows
\begin{eqnarray}
  \label{pairs}
P_{i,\mathcal{N}(i)}\left(x_i|\{x\}_j,j\in\mathcal{N}(i)\right)=\prod_{j\in\mathcal{N}(i)}P_{i,j}\left(x_i|x_j\right),
\end{eqnarray}
where, up to a normalization constant
\begin{eqnarray}
  \label{pairs1}
  P_{i,j}(x_i|x_j)\propto \exp (\beta J_{i,j} x_i x_j).
\end{eqnarray}
The parameter $\beta$ plays the role of a global amplifying factor and, although we shall eventually set $J_{i,j}=J$, we have introduced
it here to make contact with statistical mechanics, where $\beta$ represents the inverse temperature and $\beta J$ is adimensional.
We stress that, under the hypothesis that the $x_i$ are Ising variables, the form (\ref{pairs1}) is general (in other words, the full information
about $P_{i,j}$ is encoded by the single parameter $J_{i,j}$).

More in general, we allow for the presence of noise encoded by a frozen local random field $h_i$ symmetrically distributed around zero, so that Eq. (\ref{pairs1}) is replaced by
\begin{eqnarray}
  \label{pairs1RF}
  P_{i,j}(x_i|h_i;x_j)\propto \exp (\beta h_i x_i+\beta J_{i,j} x_i x_j).
\end{eqnarray}

Under our null hypothesis ($x_i\in\{-1,1\}$ and Eqs. (\ref{pairs}) and  (\ref{pairs1RF})), the MPA reads~\cite{Yedidia,Yedidia2,Mezard_Montanari}
\begin{align}
\label{BP01}
& \beta u_{i\to j}(t+1)=\tanh^{-1}\left[\right .\nonumber\\
& \left. \tanh(\beta J_{i,j})\tanh\left(\beta h_i+\sum_{k\in\partial i\setminus j}\beta u_{k\to i}(t)\right)\right],
\end{align}
where $\partial i$ stands for the set of neighbors of $i$ and
the sum involves all the neighbors but node $j$. 
Given a set of initial conditions of the messages, $\{u_{i\to j}(t=0)\}$, one can iterate Eq. (\ref{BP01}) and analyze the emerging dynamics.
In particular, one can evaluate the mean values of the variables $x_i$ at a given time $t$ by summing over all the messages arriving at site $i$ at time $t$.
The result is
\begin{eqnarray}
\label{mag}
&& x_i(t)=\media{x_i}_t=\tanh\left(\beta h_i + \sum_{k\in\partial i}\beta u_{k\to i}(t)\right),
\end{eqnarray}
where
\begin{eqnarray}
\label{mag}
&& \media{x_i}_t=\sum_{x_1,\ldots,x_N}P^{t}\left(x_1,\ldots,x_N\right)x_i.
\end{eqnarray}
Let us first focus on the case without noise: $h_i\equiv 0$. 
Particularly important are the non-trivial stationary solutions $\{u_{i\to j}\}\defi\{u_{i\to j}(t\to \infty)\}$
whose existence is guaranteed as soon as the trivial solution $\{u_{i\to j}=0\}$ becomes unstable.
Note that there might exist also non-stationary solutions. For simplicity however, in this Section we focus only
on stationary solutions. 
As a first pedagogical example, let us suppose that
$\mathcal{G}$ is a regular random graph of constant degree $k$ ($\sum_{j}{A}_{i,j}=k$) and that the couplings are all equal ($J_{i,j}\equiv J $).
With such homogeneous assumptions, the dynamics can only have homogeneous stationary solutions $u_{i,j}\equiv u$
and we can easily find the critical parameter defined as the minimal value beyond which a solution with $u\neq 0$ exists. 
From Eq. (\ref{BP01}),
by expanding for small $u$
\begin{eqnarray}
\label{BP02}
&& \beta u=\left(k-1\right)\tanh(\beta J)\beta u+\mathop{O}(u^3),
\end{eqnarray}
which implies the following critical point
\begin{eqnarray}
  \label{BP08}
  && 1=\left(k-1\right)\tanh(\beta J).
\end{eqnarray}
Above the critical point, the stationary solution undergoes a second-order phase transition or, in dynamical terms, a bifurcation takes place
where the trivial solution ceases to be stable and two opposite (symmetrical around zero) non trivial solutions emerge.
In general, these solutions correspond to the spontaneous magnetization $m_0$ of the Ising model built on the top of the graph $\mathcal{G}$,
characterized by the following Hamiltonian 
\begin{eqnarray}
  \label{Ising}
  && H= -\sum_{(i,j)\in E} J_{i,j}x_i x_j -h\sum_ix_i,
\end{eqnarray}
where $h$ is a homogeneous external field
to be sent to zero eventually. 
It is a well known property of the MPA that, if the graph $\mathcal{G}$ is locally tree-like~\cite{Review}, then
\begin{eqnarray}
  \label{Ising2}
  \lim_{N\to\infty}\media{x_i}_G=\lim_{N\to\infty}\lim_{t\to\infty} x_i(t),
\end{eqnarray}
where $\media{\cdots}_G$ stands for average over the Gibbs distribution $P_H\propto \exp(-\beta H)$.
In particular, for the spontaneous magnetization we have
\begin{eqnarray}
  \label{Ising3}
m_0=\lim_{h\to 0^+}\lim_{N\to\infty}\lim_{t\to\infty} \frac{1}{N} \sum_i x_i(t).
\end{eqnarray}
Note however that, for any finite graph, 
if $h=0$ we have $\media{x_i}_G=0$ due to the $\mathbb{Z}_2^N$ symmetry of the Hamiltonian (\ref{Ising}), unless one starts with positive initial conditions
(in fact, it is enough to consider messages with the same sign up to an order $\mathop{O}(1/N)$).
This symmetry is also reflected
in the MPA when the initial conditions are chosen to be symmetrically distributed around zero. Moreover, for exact trees (\textit{i.e.}, zero loops)
$\media{x_i}_G=0$ for any value of $\beta$.

More in general, if $\mathcal{G}$ is locally tree-like, the stationary distribution $P(x_1,\ldots,x_N)$ of the MPA reproduces
any correlation function of the associated Ising model~\cite{Mezard_Montanari}. 
Furthermore, it is also known that the MPA produces reasonable approximations for the magnetization of the model (\ref{Ising}) even in loopy graphs.
However, we do not need to make any assumption about the graph $\mathcal{G}$ other than ordinary: it is simple, undirected, sparse and connected.
When $\mathcal{G}$ has dense loops, we still assume that our dynamics is defined via the MPA, even if its connection with the Ising model (\ref{Ising}) is poor and,
more in general, even if the MPA does not converge to a stationary distribution. 

Finally, let us consider the general case with random fields $h_i$, drawn for example from a uniform distribution centered on zero. Also in this case, if the MPA converges,
the associated Ising model, known as random-field Ising model, provides the corresponding stationary distribution. Now, however, the random field prevents
the model to get a spontaneous magnetization. Rather, the random field Ising model is characterized by local magnetizations that will be crucial for
our exploration of synchronization (these local magnetizations are somehow related to an elusive spin-glass phase~\cite{RFIM_Ricci}).
The use of the random field Ising model is not new in the community-detection framework~\cite{Son,Peixoto}.
However, besides the different methodology
(here based on the MPA, while Ref.~\cite{Son} uses Monte Carlo and Ref.~\cite{Peixoto} uses Pseudolikelihood),
we use the random field Ising model for synchronization-detection, which is independent from the notion of community. 
As we have already said, the key notion for synchronization is symmetry, not community.

\section{Our algorithm}
Here we explain our general procedure for solving the target defined in Sec. \ref{Target}.
As already anticipated, for simplicity we set $J_{i,j}=J>0~ \forall (i,j)\in E$. Applications with heterogeneous couplings will be considered elsewhere.
Hereafter we also assume $\beta=1$.

For each given value of the coupling $J$, we evolve the MPA Eqs. (\ref{BP01}) with a fixed distribution of the random fields $h_i$
and given initial conditions for the $|E|$ incoming messages $u_{i\to j}$
(\textit{i.e.}, for any fixed node $j$ we consider the messages arriving from its $k$ neighbors $i_1,i_2,\ldots,i_k$). We do this
until a sufficiently large number of iterations $t=t_{\mathrm{max}}$ such that all the $2|E|$ messages reach each a stationary value within a tolerance $\epsilon$:
\begin{eqnarray}
  \label{disttime}
  |u_{i\to j}(t+1)-u_{i\to j}(t)|<\epsilon, \quad \forall (i,j) \in E.
\end{eqnarray}
If the required $t_{\mathrm{max}}$ is too large or if does not exist, we shall still use the messages collected at the largest available
time $t_{\mathrm{max}}$ as follows: we first run the MPA with a moderate trial value of $t_{\mathrm{max}}$ and evaluate the percentage of messages that have reached stationarity,
then, we run again the MPA with a lager value of $t_{\mathrm{max}}$ and evaluate the percentage of messages that have reached stationarity, and so on. When
we observe that this percentage has reached a stationary value, we choose the corresponding value of $t_{\mathrm{max}}$ for applying the MPA.

As for the $N$ independent random fields $h_i$, they are drawn from a uniform distribution with support $[-h^*,h^*]$
for some $h^*\geq 0$ (which includes the case without noise $h^*=0$).
At the end of the iteration we evaluate the $N$ mean values $x_i(t)$ of the state variables via Eqs. (\ref{mag}).
Finally, we make a partition into $Q$ clusters of these $N$ values according to Eq. (\ref{dist});
\textit{i.e.}, we say that $x_i$ and $x_j$ belong to the same cluster if
\begin{eqnarray}
  \label{distTE2}
  |x_i-x_j|<\epsilon.
\end{eqnarray}
We check Eq. (\ref{distTE2}) for all pairs of nodes $(i,j)$, next, for finding the related clusters
we simply order the $N$ values $x_i$ 
in increasing order and finally group them into the $Q$ clusters using again Eq. (\ref{distTE2}).

Of course, everything makes sense if the $\epsilon$ chosen in Eq. (\ref{distTE2}) is close
to the $\epsilon$ chosen in Eq. (\ref{disttime}). As a rule of thumb, we simply chose the two equal with each other and equal
to the best available machine precision of the specific computer where the MPA is supposed to be run. Typically,
for ordinary computers we have $\epsilon=10^{-15}$ or $\epsilon=10^{-18}$, but finer choices are also possible
(at some memory cost).
In Appendix \ref{AppA} we show a diagrammatic synthetic description of our algorithm.

We specialize the algorithm with three variants:
with uniform initial conditions, with random initial conditions, and with noise.
Each case is equipped
with a simulation study of a benchmark synthetic graph whose description
is reported in the caption of Fig.~\ref{figQ207}.
Below we discuss each version in detail. 

\subsection{Noiseless version with positive initial conditions}
A dynamics with positive initial conditions—homogeneous or not—is not intended to model realistic systems.
Rather, we employ this regime as a diagnostic tool to probe the stable symmetries of the underlying graph $\mathcal{G}$.
For a given value of the coupling 
$J$, our goal is to identify a partition of $\mathcal{G}$
into $Q$ clusters of stable synchronized nodes and to interpret these clusters in terms of TE groups of $\mathcal{G}$.
To this end, we assume positive and homogeneous initial conditions and focus exclusively on stationary solutions
(non-homogeneous initial conditions lead to identical results, provided they remain non-negative).

Let us consider the associated Ising model (\ref{Ising}). If the mean degree 
$\media{k}$ of $\mathcal{G}$ is not too small (typically $\media{k}>1$ suffices; see Eq.~(\ref{BP08}))),
there exists a finite critical coupling $J_1=J_c$ above which the trivial stationary solution $u_{i\to j}\equiv 0$ becomes unstable.
At this point, the $\mathbb{Z}^N_2$ symmetry is broken and a stationary global magnetization 
$m$ emerges, with $m=m_0$ (up to $\epsilon$-corrections),
where $m_0$ is the spontaneous magnetization (\ref{Ising3}).
In physical terms, a finite fraction of spins aligns to a common value.

From the synchronization perspective, this instability signals the formation of clusters of nodes sharing identical states, which can be discriminated via their local mean magnetization. Typically (but not exclusively), the first clusters to form correspond to regions of higher connectivity. As 
$J$ is further increased, existing synchronized clusters smoothly increase their magnetization, while additional clusters emerge at distinct critical points.
Each such event corresponds to a non-analytic behavior of the order parameter, more precisely to a discontinuity in
$d m_0(J)/d J$. We denote the resulting sequence of critical couplings by $\{J_1,J_2,\ldots\}$.
Eventually, for sufficiently large $J$, different clusters merge and a saturation regime is reached.

Even under the assumption of positive initial conditions, the 
$Q$ synchronized clusters do not generally coincide with the 
$Q_{\mathrm{TE}}$ groups of TE nodes of $\mathcal{G}$.
in general, the $l$-th cluster contains portions of $n^{(l)}\geq 1$ distinct TE groups, with associated fractions $0\leq \alpha_1,\ldots,\alpha_{n^{(l)}}\leq 1$.
Nevertheless, as $J$ increases, most of these fractions tend toward either 0 or 1, so that each cluster eventually contains entire TE groups.
In particular, we expect that TE groups whose internal degree is not smaller than their external degree, $k^{l;\mathrm{OUT}}\leq k^{l;\mathrm{IN}}$,
are more likely to be fully embedded within a synchronized cluster.

\begin{figure}[htb]
    \centering
    \includegraphics[width=0.8\columnwidth,clip]{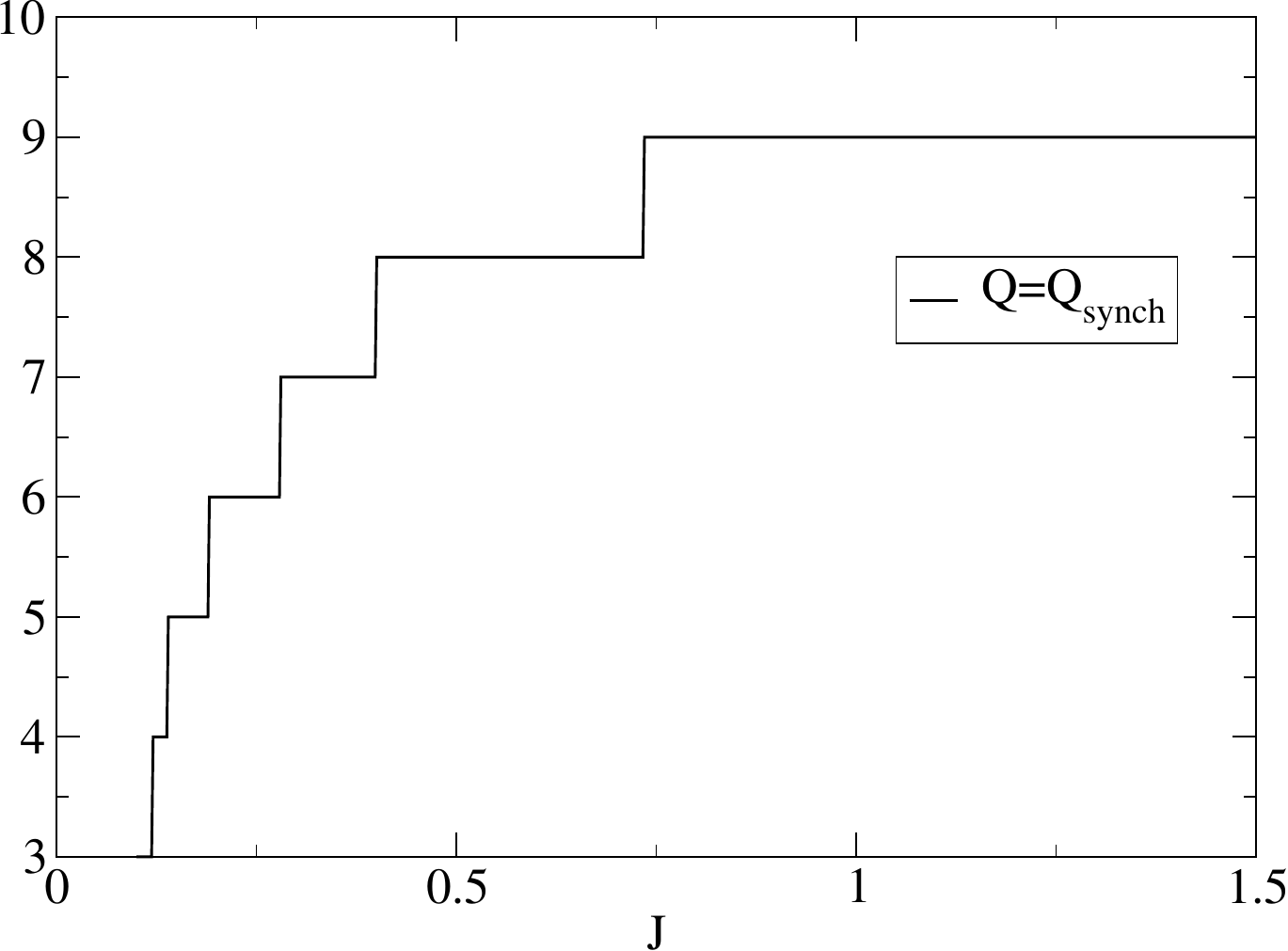}
      \caption{
        Number of synchronized clusters $Q$ as a function of the coupling $J$ with $\beta=1$
        obtained by evolving the noiseless MPA with positive initial conditions in a synthetic graph $\mathcal{G}$ with $N=744$ nodes
        and mean degree $\mediaT{k}=3.32 \pm 1.84$. The graph owns $Q_{\mathrm{TE}}=207$ groups of TE nodes with size between 2 and 5.
        Among these groups, 122 are characterized for having internal degree not smaller than the external one. The MPA has been realized with $t_{\mathrm{max}}=10^4$
        (which is enough for having stationarity within precision $\epsilon=10^{-18}$ for $J\in[0,2]$).
        In this example, the participation ratio $p$, defined as the total number of synchronized nodes
        divided by the total number of nodes, is identically equal to 1 (which amounts to $Q=Q_{\mathrm{synch}}$, $Q_{\mathrm{synch}}$
        being the number of clusters with size larger than 1).}
      \label{figQ207}
  \end{figure}

\begin{figure}[htb]
    \centering
    \includegraphics[width=0.8\columnwidth,clip]{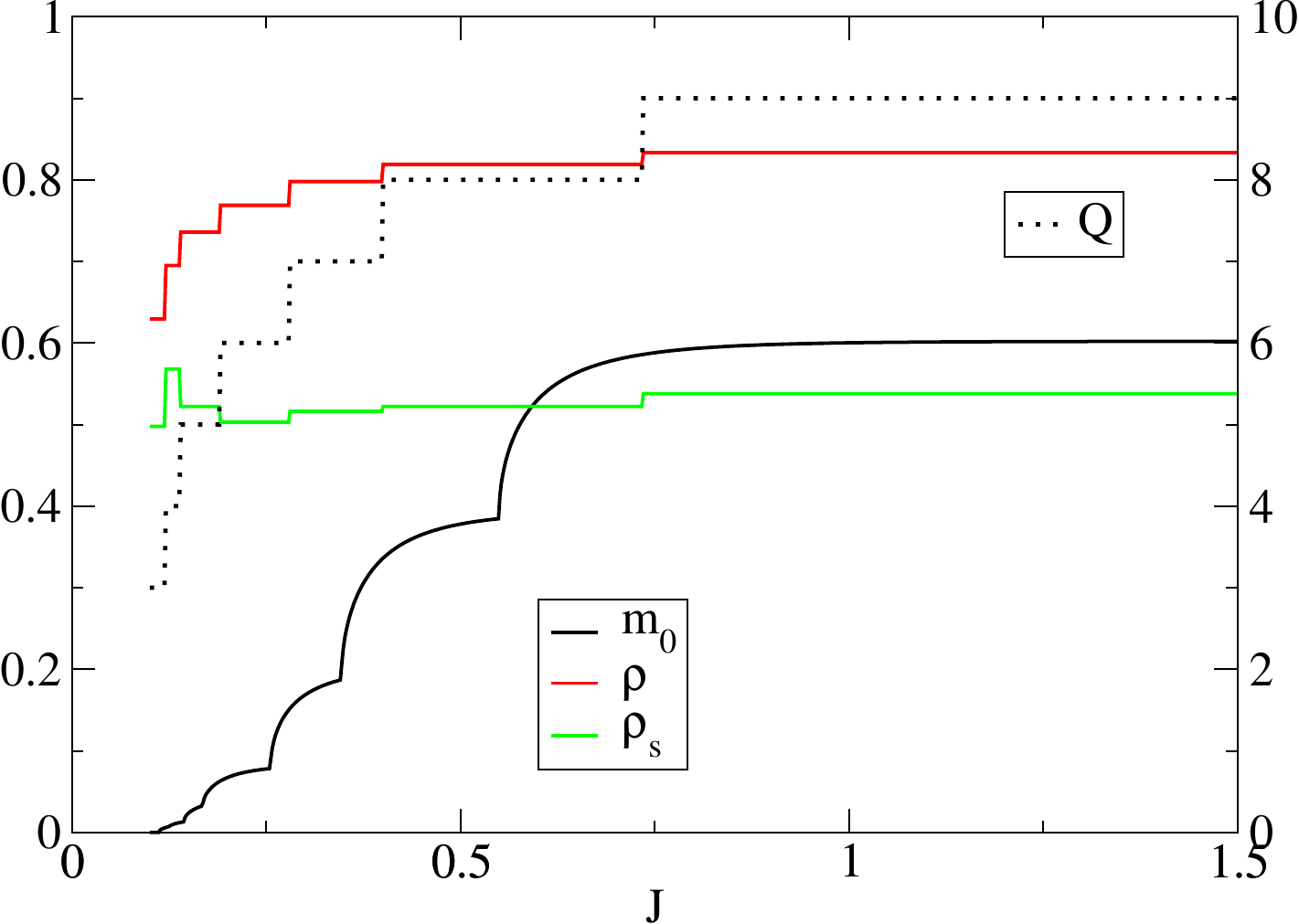}
      \caption{
        Spontaneous magnetization $m_0$ obtained by evolving the noiseless MPA with positive initial conditions in the synthetic graph $\mathcal{G}$ of
        Fig.~\ref{figQ207}. The function $\rho$ provides the percentage of groups of TE nodes fitting entirely within one of the $Q$ synchronized clusters.
        Similarly, $\rho_s$ provides the percentage of those groups of TE nodes
        characterized for having internal degree not smaller than the external one. We report also the same $Q(J)$ of Fig.~\ref{figQ207}
        (dotted plot with scale on the right axis) for better highlighting
        the correspondence with $m_0(J)$.}
        \label{figmagQ207}
  \end{figure}

We have verified this hypothesis by comparing the 
$Q$ synchronized clusters obtained from the dynamics with the $Q_{\mathrm{TE}}$
TE groups in synthetically generated networks. An illustrative example with $Q_{\mathrm{TE}}=207$ is reported in
in Figs. \ref{figQ207} and \ref{figmagQ207}.
These results highlight the role of TE groups as effective nucleation points for synchronization,
in agreement with previous findings obtained within Kuramoto-like dynamics~\cite{OstilliTE}.

\subsection{Noiseless version with random initial conditions}
We consider this version when a more realistic dynamics is required, while still neglecting the presence of noise.
As a matter of fact, even in the present more general case, the description of cluster of synchronized nodes in terms of groups of TE nodes remains partially valid.
More precisely, it remains valid for values of $J$ far from points where violent oscillations of $Q$ occur on varying $J$.
Indeed, when random initial conditions are allowed,
the dynamics becomes significantly richer, giving rise to a complex pattern of alternating synchronization and desynchronization regimes.
In this case, synchronization intervals are interrupted by very abrupt desynchronization events occurring when 
$J$ approaches one of the critical couplings $\{J_1,J_2,\ldots\}$
introduced in the previous version. This behavior can be conveniently characterized by monitoring both the total number of clusters 
$Q$ and the number of synchronized clusters $Q_{\mathrm{synch}}$
as functions of $J$. Here, whereas
$Q$ denotes the total number of clusters (including single-node clusters), $Q_{\mathrm{synch}}$ counts only clusters of size larger than one.

Away from critical points, one typically finds $Q_{\mathrm{synch}}\simeq Q$, indicating that most nodes belong to synchronized groups.
As $J$ approaches a critical value $J_c\in\{J_1,J_2,\ldots\}$ from below, both 
$Q$ and $Q_{\mathrm{synch}}$ initially increase rapidly, signaling a progressive fragmentation of the network into smaller clusters.
Closer to the critical point, while $Q$ continues to grow approximately exponentially,
$Q_{\mathrm{synch}}$ remains nearly constant and much smaller than $Q$.
In this regime, the difference $Q-Q_{\mathrm{synch}}$
corresponds to a large number of isolated nodes, \textit{i.e.}, nodes that do not synchronize with any other node.

\begin{figure}[htb]
    \centering
    \includegraphics[width=0.8\columnwidth,clip]{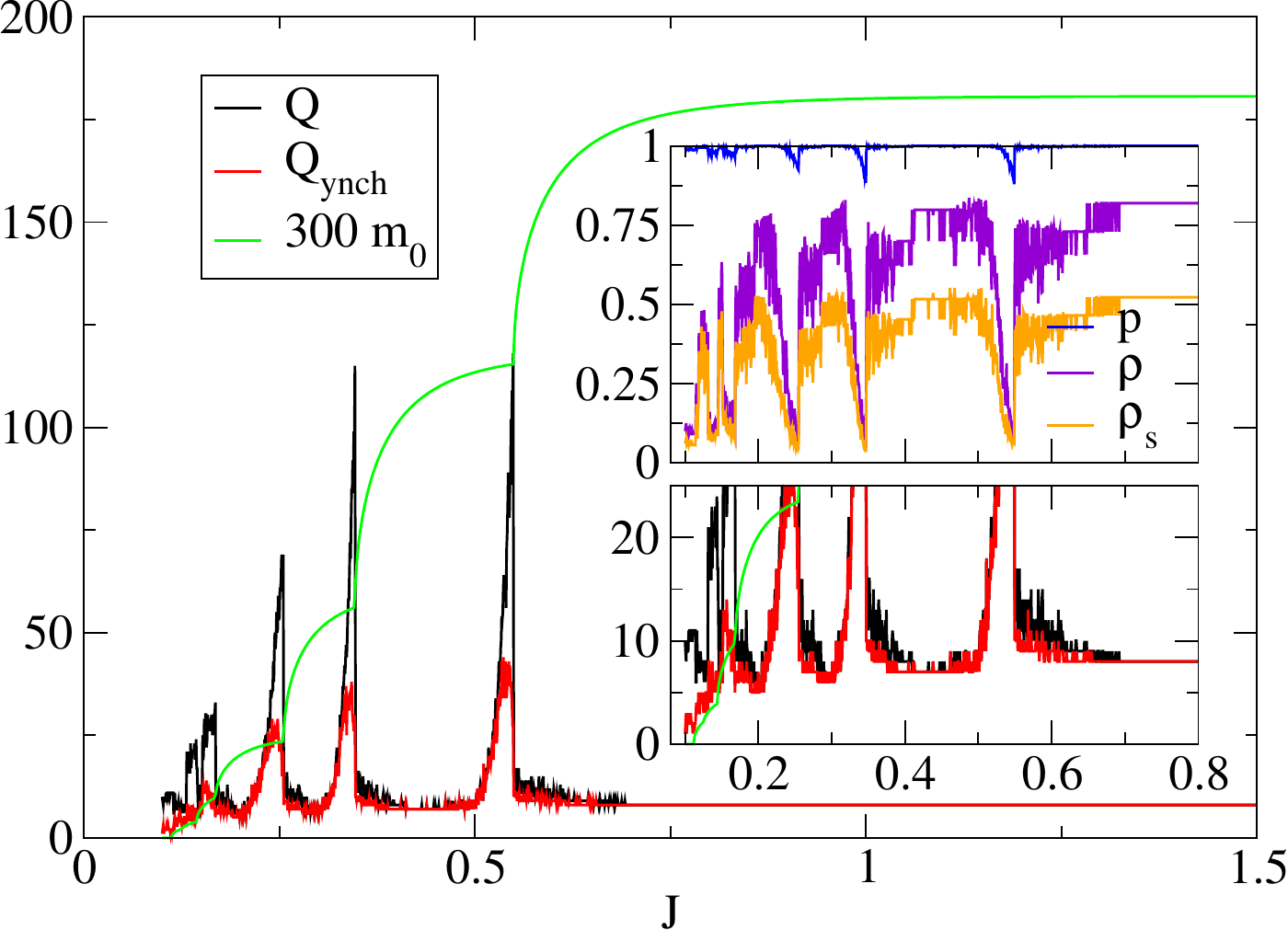}
    \caption{
      The number $Q$ of clusters and the number $Q_{\mathrm{synch}}$ of clusters having size larger than 1 as a function of $J$ obtained
      via the MPA with random initial conditions applied to the same graph of Fig.~\ref{figQ207}.
      The MPA has been realized with $t_{\mathrm{max}}=5\times 10^5$
        (which is enough for having stationarity within precision $\epsilon=10^{-18}$ for $J\in[0,1.5]$).
      The green plot represents the spontaneous magnetization (the one obtained for Fig. \ref{figmagQ207}) magnified by a factor 300.
      It shows that the explosive desynchronization points correspond to the critical points $J_1,J_2,\ldots$ of the spontaneous magnetization $m_0(J)$.
      The bottom Inset shows more details while the top Inset shows: the participation ratio $p$ (defined as the total number of synchronized nodes
      divided by the total number of nodes), and the quantities $\rho$ and $\rho_s$ providing the percentages about the groups of TE nodes
      fitting entirely within one of the $Q$ synchronized clusters (as defined in caption of Fig.~\ref{figmagQ207}).}
    \label{figmagrandomQ207}
  \end{figure}

Exactly at $J=J_c$, both $Q$ and $Q_{\mathrm{synch}}$
undergo a discontinuous drop to moderate values. For $J>J_c$, the system re-enters a smoother synchronization regime until the next critical point is reached,
where the same pattern repeats. An illustrative example of this behavior, obtained by applying the algorithm to the synthetic network discussed in Fig.~\ref{figQ207},
is shown in Fig.~\ref{figmagrandomQ207}.

  \subsection{Noisy version}
\begin{figure}[htb]
    \centering
    \includegraphics[width=0.8\columnwidth,clip]{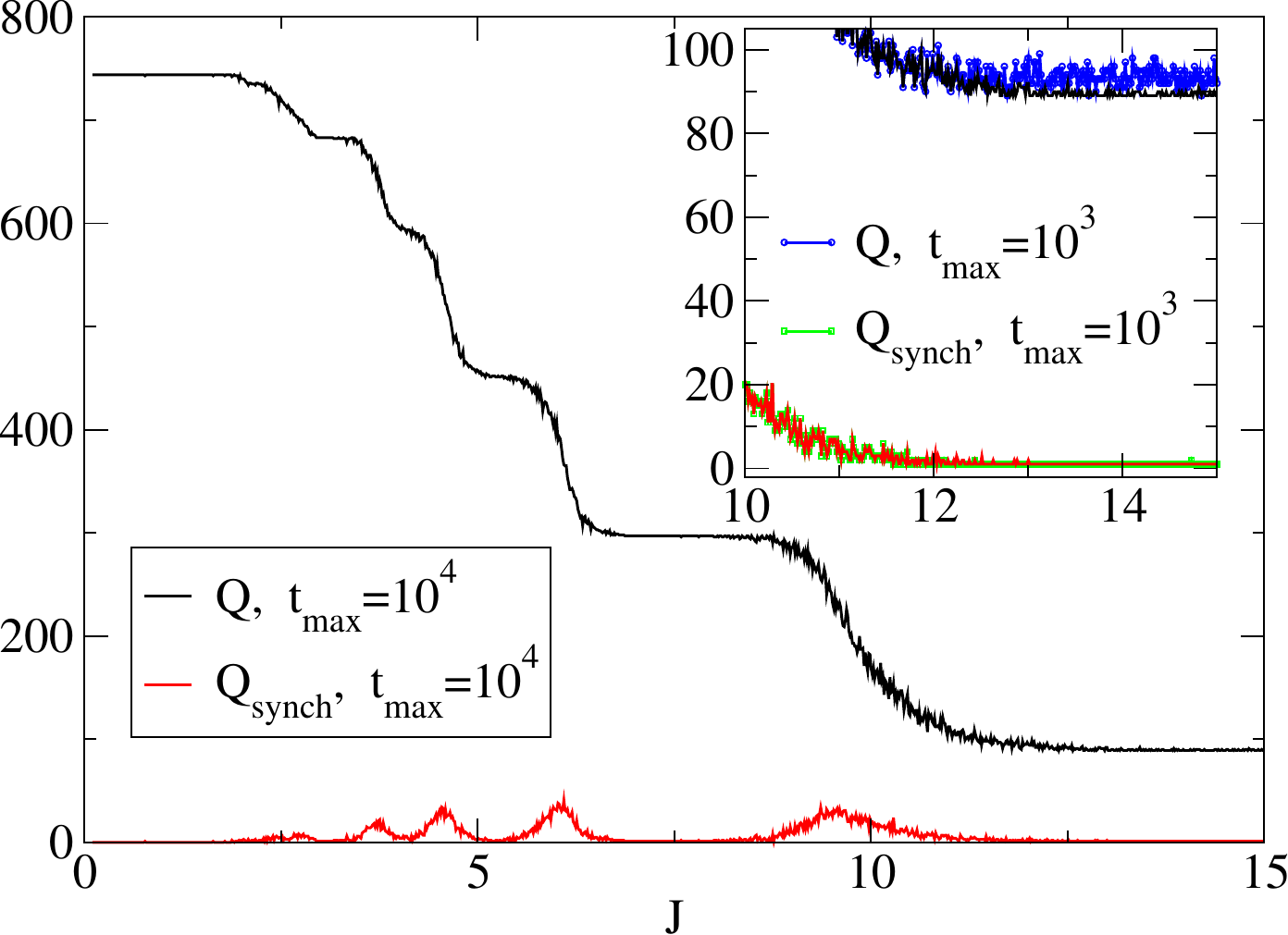}
    \caption{
      The number $Q$ of clusters and the number $Q_{\mathrm{synch}}$ of clusters having size larger than 1 as a function of $J$ obtained
      via the MPA with random initial conditions and random external field applied to the same graph of Fig.~\ref{figQ207}.
      The random external fields have been drawn independently from the interval $[-1,1]$ for each node and from each value of $J$.
      Here, the MPA has been realized by using two different values of $t_{\mathrm{max}}$:
      $t_{\mathrm{max}}=10^3$ (with about $9 \%$ of messages becoming stationary) and
      $t_{\mathrm{max}}=10^4$ (with about $35 \%$ of messages becoming stationary).
      The corresponding plots of $Q$ and $Q_{\mathrm{synch}}$ are essentially equal on the scale of the figure, while they show small visible differences
      in the Inset only for large values of $J$. The first plateau occurring for $J\in[0,1.7\pm 0.1]$ corresponds to $Q=N=744$, \textit{i.e.}, no synchronization.}
    \label{figmagnoisyQ207}
  \end{figure}

\begin{figure}[htb]
    \centering
    \includegraphics[width=0.8\columnwidth,clip]{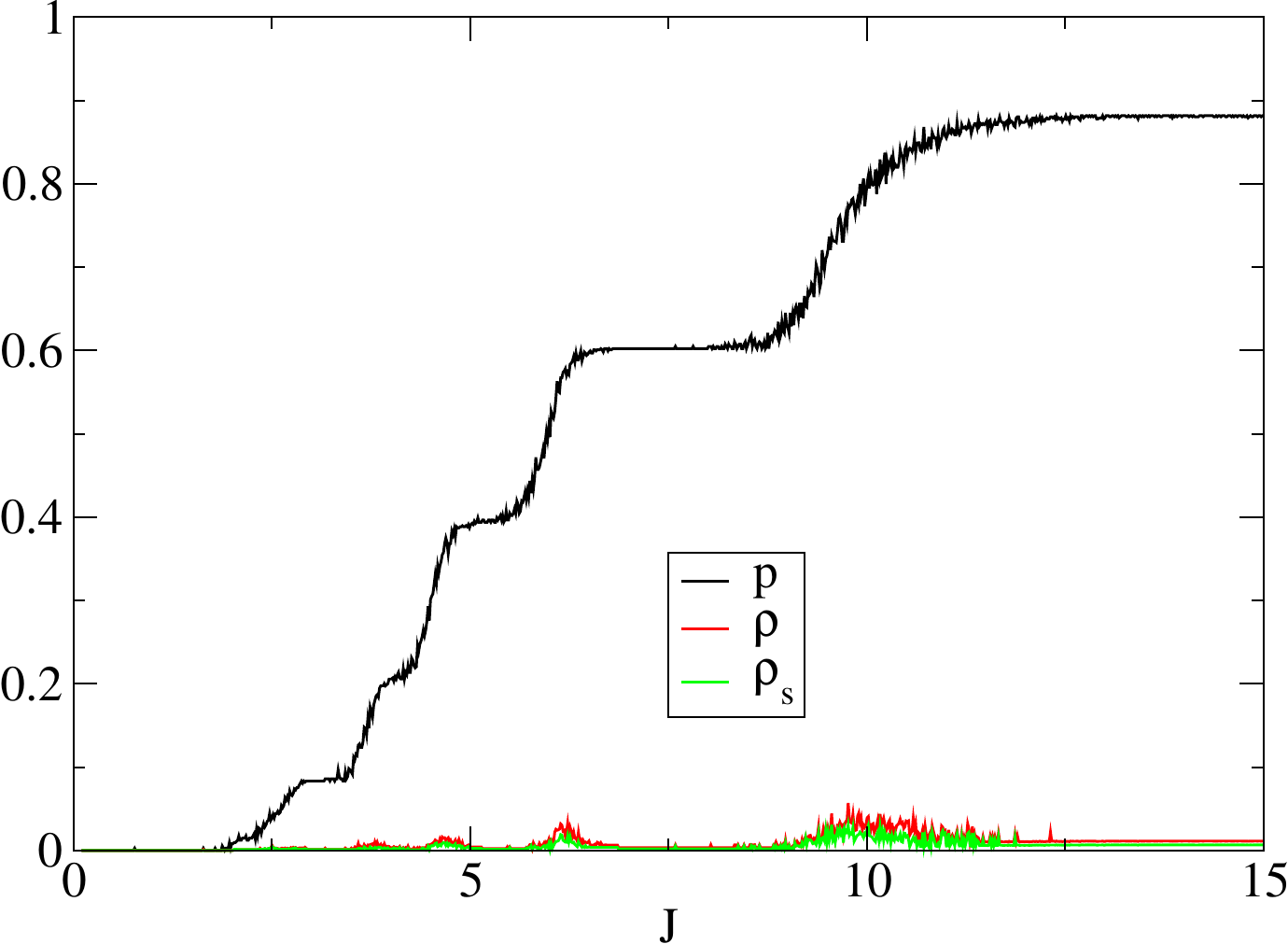}
    \caption{
      The participation ratio $p$ (defined as the total number of synchronized nodes
      divided by the total number of nodes), and the quantities $\rho$ and $\rho_s$ providing the percentages about the groups of TE nodes
      (as defined in caption of Fig.~\ref{figmagQ207}) from the same MPA used for Fig.~\ref{figmagnoisyQ207}.
      The plots correspond to the data obtained by using $t_{\mathrm{max}}=10^4$, while the plots corresponding to the data obtained by using $t_{\mathrm{max}}=10^3$
      show no visible difference on this scale.}
    \label{figmagnoisy2Q207}
  \end{figure}

\begin{figure}[htb]
    \centering
    \includegraphics[width=0.8\columnwidth,clip]{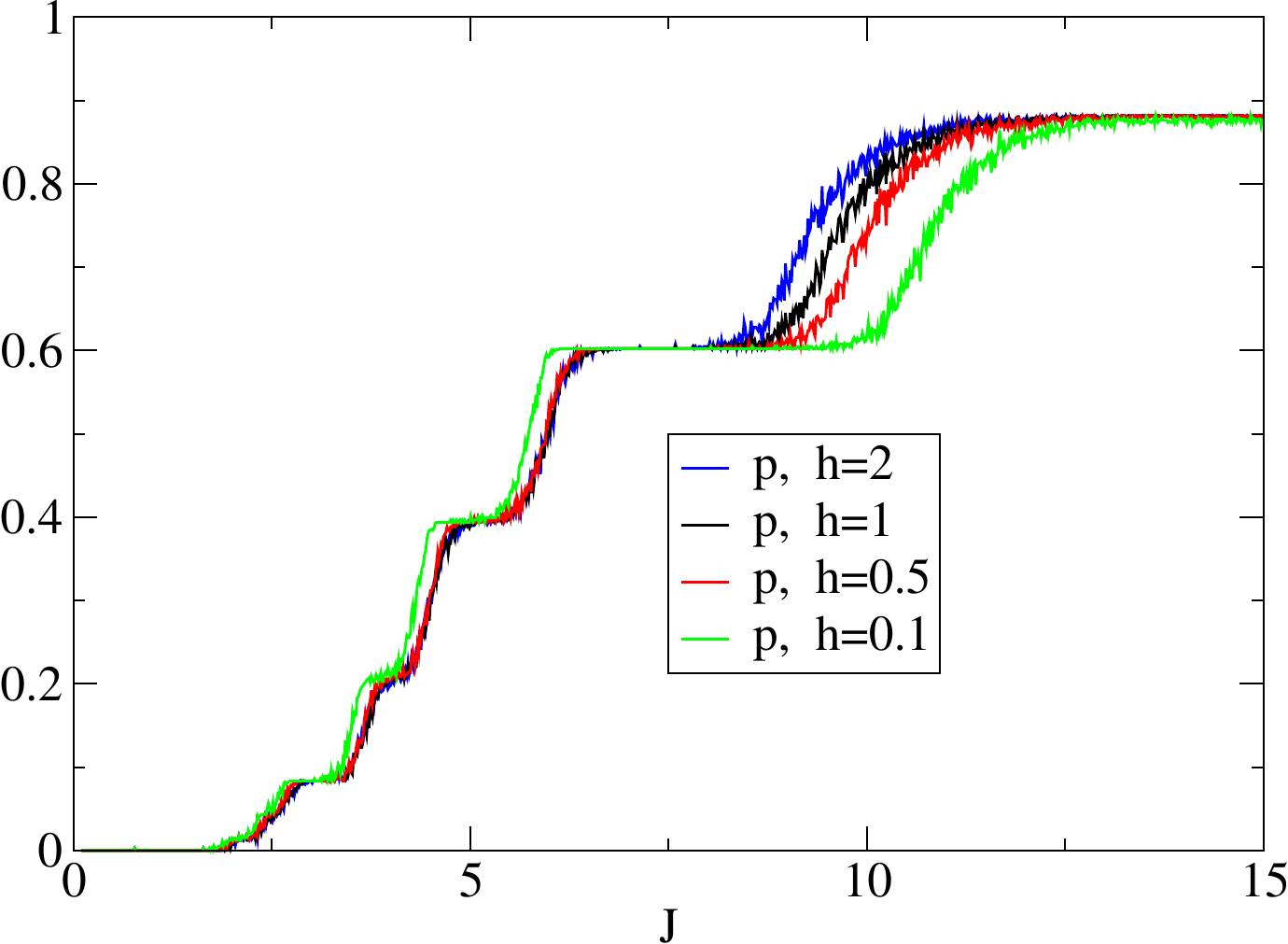}
    \caption{
      Comparing the participation ratio $p$ for four different values of $h^*$ (the amplitude  of the random field)
      obtained from the MPA with random initial conditions and $t_{\mathrm{max}}=10^4$ applied
      to the same synthetic graph discussed in Fig.~\ref{figQ207}. Despite the percentage
      of stationary messages depends on $h^*$ (ranging between $10\%$ and $35\%$ for these values of $h^*$),
      we have verified that increasing $t_{\mathrm{max}}=10^4$ implies very little changes which are not visible on the scale of the present graph.
      Notice that the number of plateaus as well as their values is independent from $h^*$.
      }
    \label{figmagnoisy3Q207}
  \end{figure}

  In the noisy version, besides using random initial conditions, we include an external random field drawn from a uniform distribution in the interval $[-h^*,h^*]$.
  It turns out that, for any $h^*>0$, the synchronization-desynchronization scenarios previously seen is completely destroyed in favor of a non decreasing behavior
  of the participation ratio $p(J)$ enriched for the presence of several plateaus, the first one of which corresponds to $p(J)=0$, \textit{i.e.}, $Q=N$.  
  As expected, the presence of noise makes also the notion of groups of TE nodes quite irrelevant.
  In Figs. \ref{figmagnoisyQ207} and \ref{figmagnoisy2Q207} we report the application of this algorithm to the same synthetic graph discussed in Fig.~\ref{figQ207}.
  We observe that, now, the global stationarity condition cannot be achieved. In other words, the presence of the external random field has the combined effect
  to make irrelevant the graph symmetries and to produce an actual oscillatory dynamics.
  Quite interestingly, as Fig.~\ref{figmagnoisy3Q207} shows, the number and the values of the several plateaus do not change by changing $h^*$. Indeed,
  also these two features seem to be uniquely related to the structure of the spontaneous magnetization $m_0$. This universal property makes this version
  of the MPA an ideal robust candidate as an unbiased probe for synchronization. 
  In conclusion, the presence of noise leads to a proper 
  synchronization pattern with a non-decreasing participation ratio and with universal features 
  where the larger the coupling the stronger the synchronization, until some saturation sets in.
  However, such a regular synchronization scenario, when compared to the case without random field, occurs at the cost of larger values of the coupling $J$.
  
\subsection{\label{AppB}Detailed partition output}
\begin{figure}[htb]
    \centering
    \includegraphics[width=0.5\columnwidth,clip]{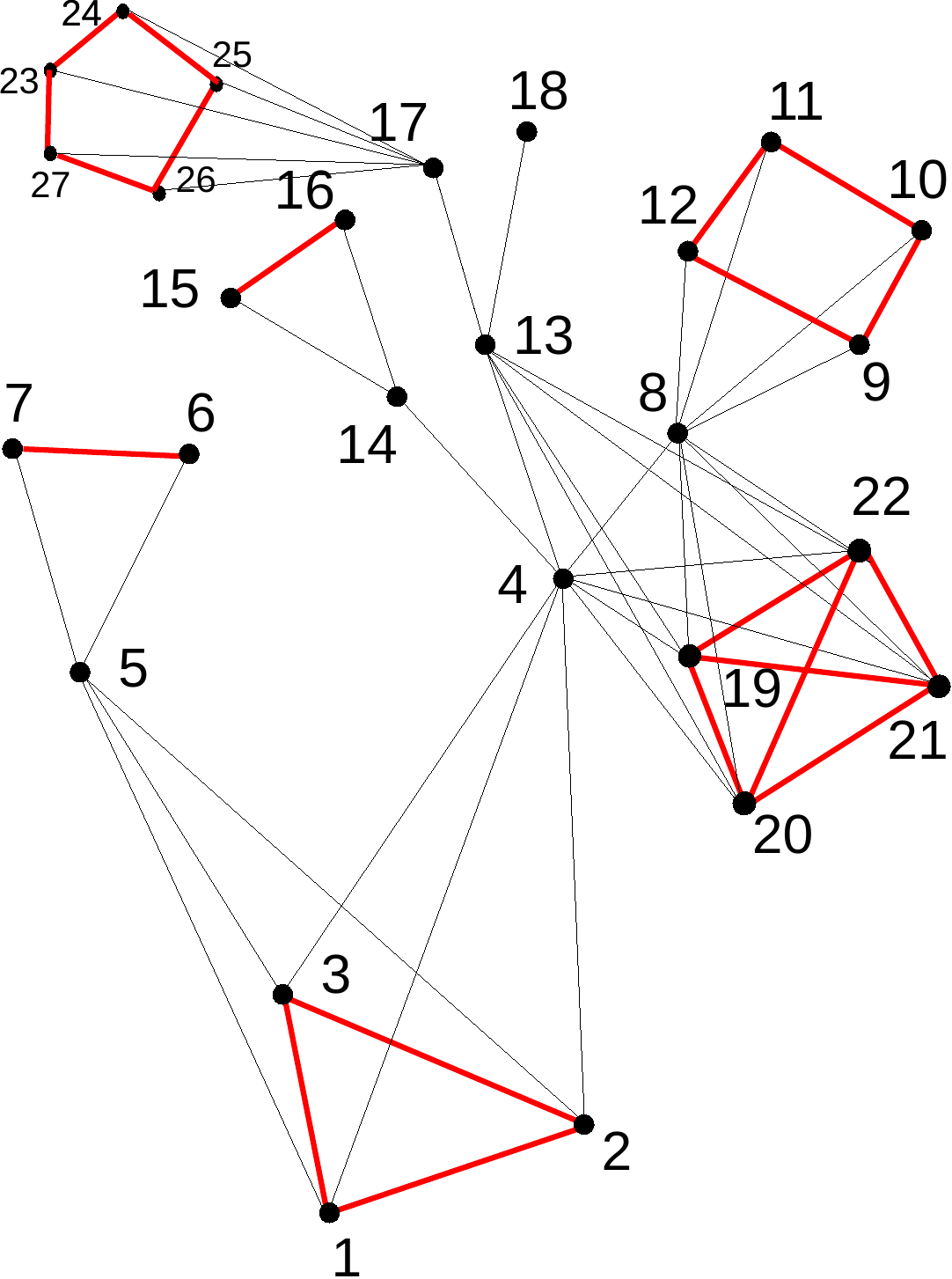}
    \caption{A toy graph with $N=27$
      nodes and $L=56$ links. The graph contains six groups of TE nodes (those having red links).
      In terms of $k^{(\mathrm{OUT})}$ and $k^{(\mathrm{IN})}$ we have:
      one group with $k^{(\mathrm{OUT})}>k^{(\mathrm{IN})}$;
      four groups with $k^{(\mathrm{OUT})}=k^{(\mathrm{IN})}$;
      and one group with $k^{(\mathrm{OUT})}<k^{(\mathrm{IN})}$.
    }
    \label{Toy_Max_2_graph}
  \end{figure}

\begin{figure}[htb]
    \centering
    \includegraphics[width=0.8\columnwidth,clip]{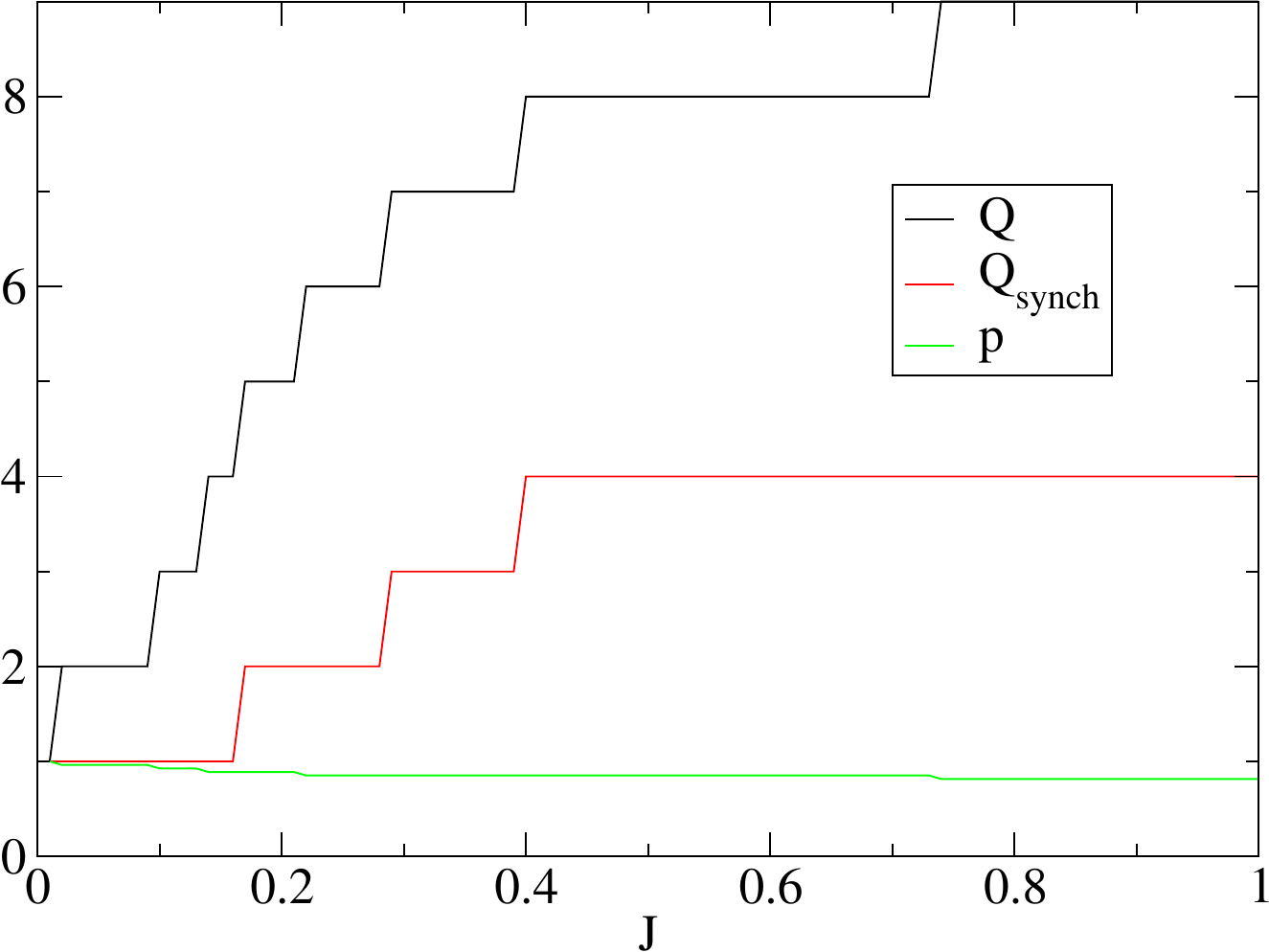}
    \caption{The number $Q$ of clusters, the number $Q_{\mathrm{synch}}$ of clusters having size larger than 1, and the participation ratio as functions of $J$ obtained
      via the MPA with uniform initial conditions applied to the graph of Fig.~\ref{Toy_Max_2_graph}.
      The MPA has been realized with $t_{\mathrm{max}}=10^4$
        (which is enough for having stationarity within precision $\epsilon=10^{-18}$ for $J\in[0,1.5]$).
    }
    \label{Toy_Max_2}
  \end{figure}

\begin{figure}[htb]
    \centering
    \includegraphics[width=0.94\columnwidth,clip]{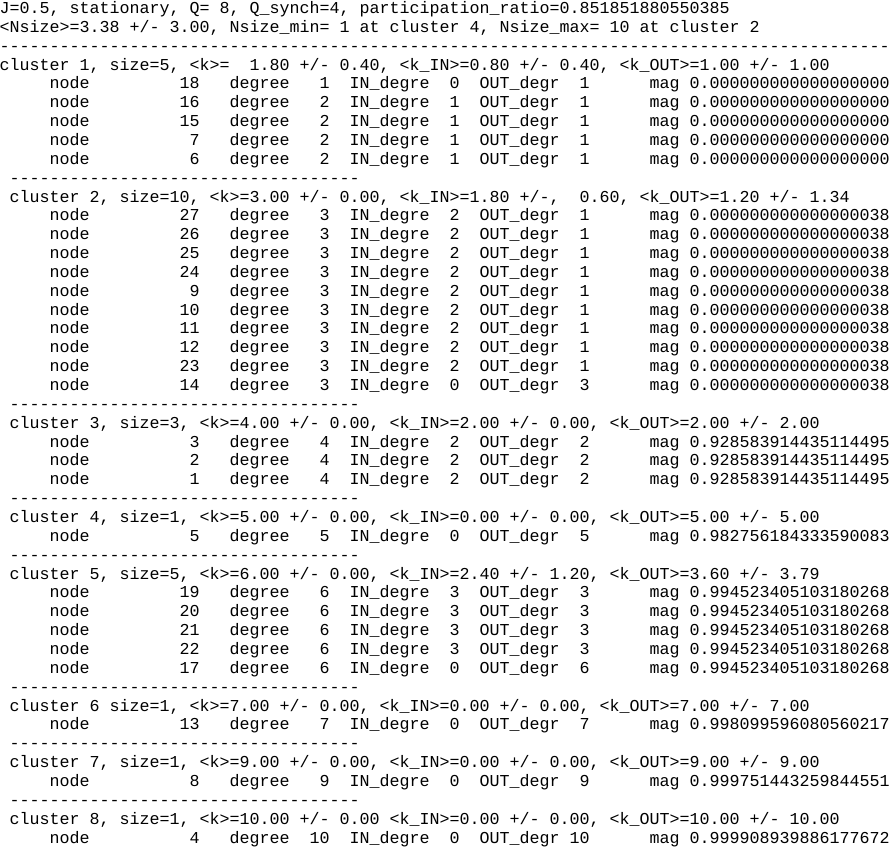}
    \caption{Detailed partition output resulting from the application of our algorithm to the graph of Fig.~\ref{Toy_Max_2_graph} for $J=0.5$.
      The first line reports the value of $J$, whether it is stationary or not, $Q$, $Q_{\mathrm{synch}}$ and $p$. The second line reports
      the mean size of the clusters, as well as their minimal and maximal values.
      The third line reports the cluster index, 1, its size, its mean degree, as well as its mean incoming and mean outgoing degrees.
      The 5 lines below this provide which nodes belong to this cluster with index 1 and, for each of them, their degree, incoming and outgoing degrees, and local magnetization.
      Similarly for the other clusters with index 2 to 8. 
    }
    \label{partition}
  \end{figure}

The partition output of our algorithm when applied to the modest graph considered before ($N=744$) is already excessively large to be print.
Here, we report the detailed partition for the much smaller graph depicted in Fig.~\ref{Toy_Max_2_graph} ($N=27$)
to show how it works.
We focus on the first version of the algorithm to see the relation
between clusters of synchronized nodes and TE groups.
From Fig.~\ref{Toy_Max_2} we see that stable values of $Q=8$ and $Q_{\mathrm{synch}}=4$
are reached for example in the interval $J\in[0.5,0.7]$. In correspondence
of these values of $J$ we get the partition output reported in Fig.~\ref{partition}.
As expected, each of the six groups of TE nodes of the graph fit entirely within one single cluster of synchronized nodes.
Note that, in this small example, the 8 clusters corresponds to 8 different degrees, which are also homogeneous
within 7 clusters. However, such a partition into different degrees is related to the small size of the graph.
For large graphs, in general, there exist several distinct clusters with equal degree and/or clusters with inhomogeneous degree,
as we have verified for example in the US power-grid. 

  \section{Application to real-world networks}
We applied the proposed algorithm to two real-world networks: the US power grid and the WordNet lexical network. With sizes 
$N=4941$ and $N=145994$, respectively, these systems do not yet fall into the Big Data regime; nevertheless, they are sufficiently large—especially the latter—to
provide a meaningful test of the scalability and efficiency of the method. In both cases, we successfully obtained a partition of the networks into synchronized groups and stored the resulting data within approximately 12 hours (about 2 seconds and 9 seconds, respectively, for each coupling value) using a single CPU core of a desktop computer (AMD Ryzen 9 3900).

\subsection{US power grid}
The US power-grid network~\cite{USpower0,USpower} consists of 
$N=4941$ nodes and has mean degree $\mediaT{k}=2.67\pm 1.79$.
Its topology represents a compromise between a geographically constrained regular graph and a random graph; in fact it is a ``small-world'' graph~\cite{USpower0,Review}.

\begin{figure}[htb]
    \centering
    \includegraphics[width=0.8\columnwidth,clip]{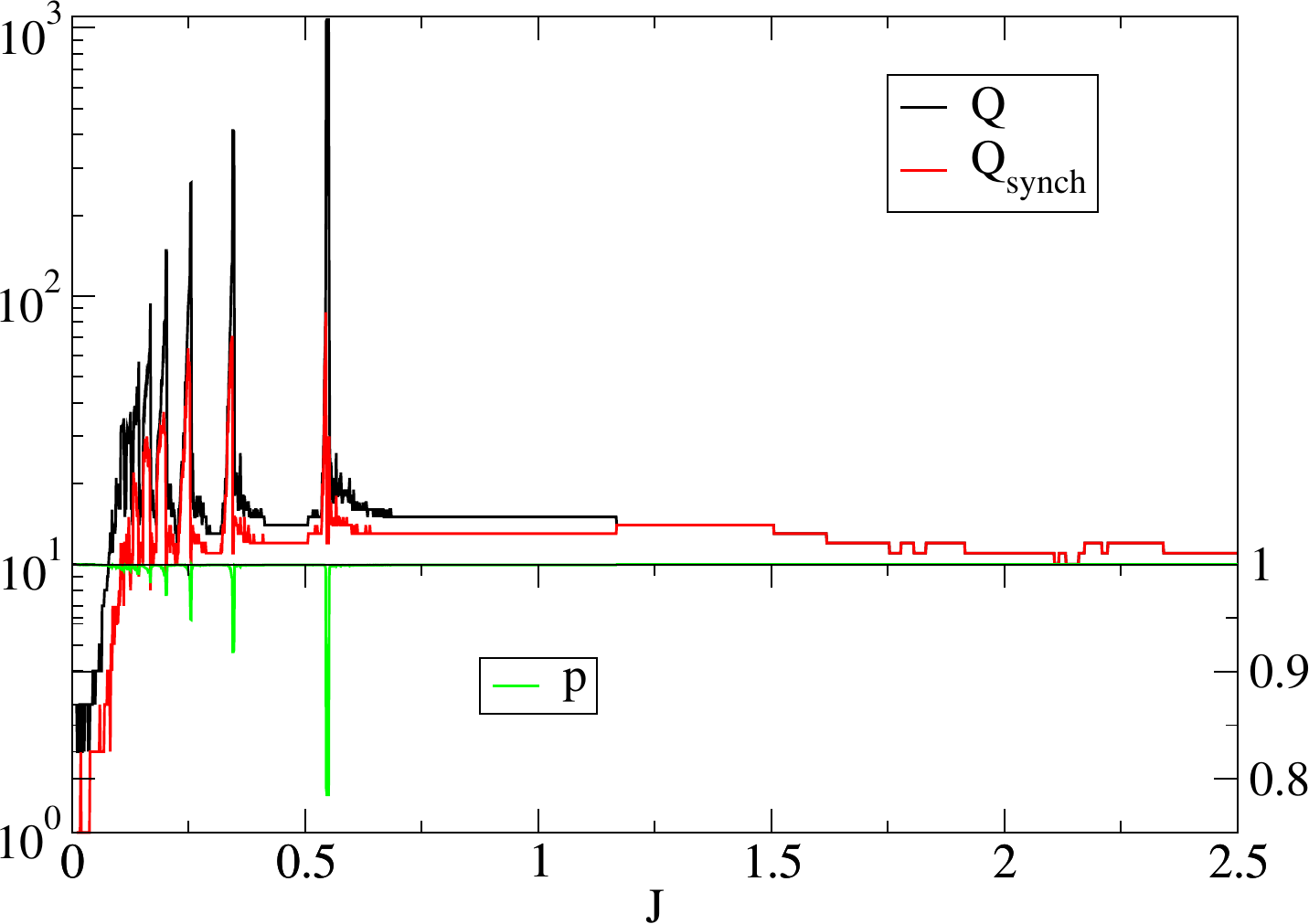}
    \caption{
      The number $Q$ of clusters, the number $Q_{\mathrm{synch}}$ of clusters having size larger than 1, and the participation ratio $p$ (scale on the right axis)
      as functions of $J$ obtained
      via the noiseless MPA with random initial conditions applied to the network of the US power grid. 
    Here we used $t_{\mathrm{max}}=5\cdot 10^3$. Peaks correspond to abrupt desynchronizations.}
    \label{power_grid}
  \end{figure}

In the noiseless case, we applied the MPA with random initial conditions. We found that choosing 
$t_{\mathrm{max}}=5\cdot 10^3$ 
iterations is sufficient to reach stationarity for approximately
$65\%$ of the messages
for$J\in[0,2.5]$; larger value of $t_{\mathrm{max}}$ yield identical results.
In Fig.~\ref{power_grid} we report the resulting behavior of $Q$, $Q_{\mathrm{synch}}$ and $p$
as functions of $J$.

We emphasize that the MPA dynamics is an artificial probe and bears no relation to the actual physical dynamics of power-grid systems, which involve generators, transformers, and continuous-time control mechanisms. It is therefore not surprising that the synchronization scenario obtained within our framework differs significantly from that obtained by simulating the true physical dynamics~\cite{Boccaletti3}. In particular, the maximal number of synchronized clusters differs substantially, with 
$Q_{\mathrm{synch}}\simeq 90$ in our case, compared with $Q_{\mathrm{synch}}\simeq 390$ reported for the physical dynamics.
However, based on our previous discussion of topologically equivalent (TE) groups as nucleation points for synchronization, we expect that each of the smaller clusters observed in the physical dynamics is entirely contained within one of the larger clusters identified by the MPA.

\begin{figure}[htb]
    \centering
    \includegraphics[width=0.8\columnwidth,clip]{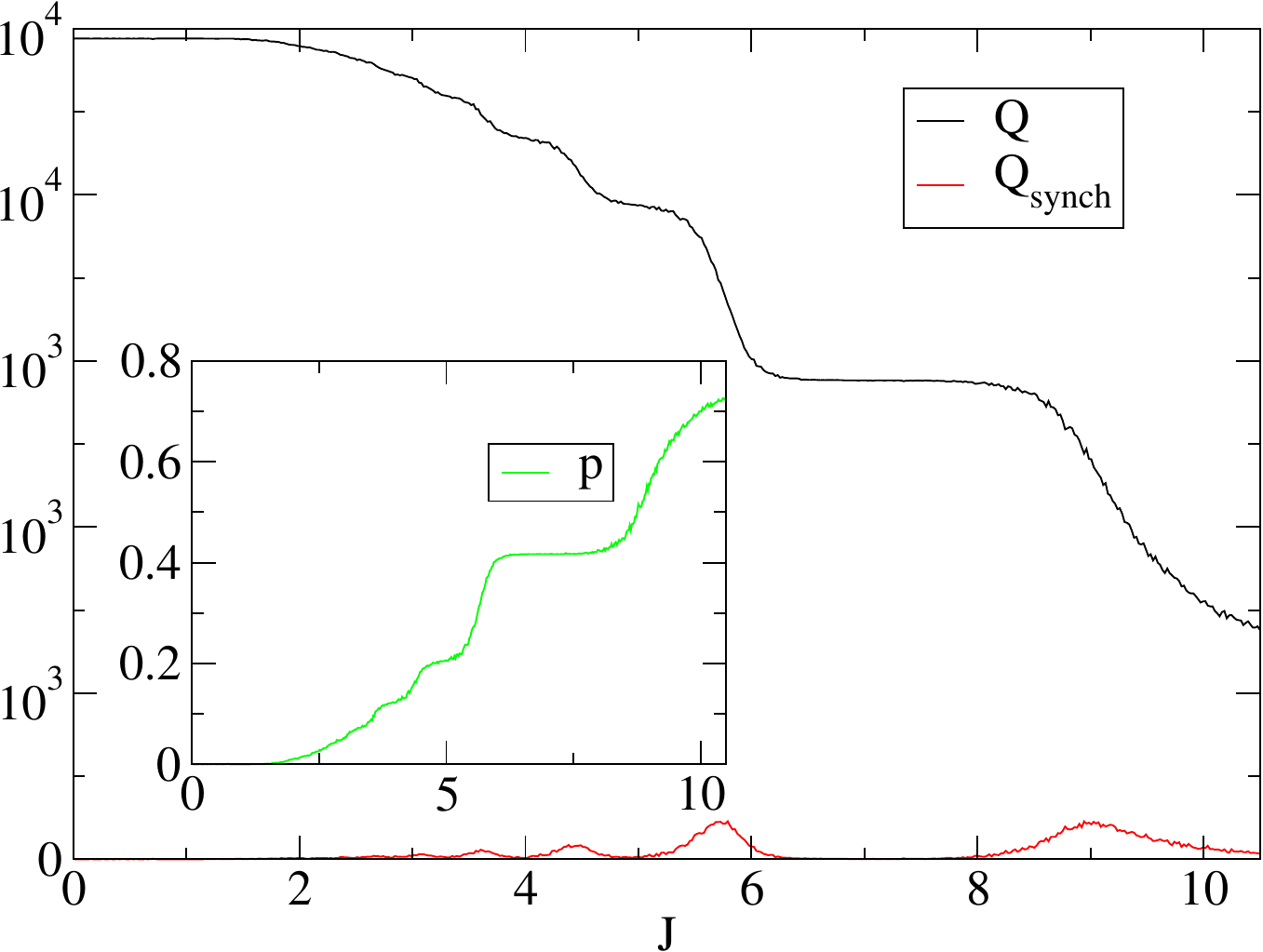}
    \caption{
      The number $Q$ of clusters, the number $Q_{\mathrm{synch}}$ of clusters having size larger than 1, and the participation ratio $p$ (Inset)
      as functions of $J$ obtained
      via the MPA with random initial conditions and random noise in the range $[-1,1]$ applied to the network of the US power grid. 
    Here we used $t_{\mathrm{max}}=5\cdot 10^4$.}
    \label{power_grid_noisy}
  \end{figure}

We then applied the MPA in the presence of noise, using 
$t_{\mathrm{max}}=5\cdot 10^4$.
The corresponding results for $Q$, $Q_{\mathrm{synch}}$ and $p$ (Inset)
are shown in Fig. \ref{power_grid_noisy}.
As expected, the inclusion of noise suppresses abrupt desynchronization events, leading to monotonic behaviors of
$Q$ and $p$ as functions of $J$, characterized by smooth plateaus.

\subsection{WordNet}
The WordNet network~\cite{Wordnet} consists of $N=145994$ nodes {with} mean degree $\mediaT{k}=3.21\pm 11.97$. 
The large variance reflects the heavy-tailed nature of the degree distribution, which is known to follow a power law.
WordNet is a directed semantic network whose nodes represent synsets (sets of synonymous words encoding concepts)
and whose edges represent lexical–semantic relations. In the present analysis, we treat the network as undirected by neglecting edge directions and removing duplicated links.

\begin{figure}[htb]
    \centering
    \includegraphics[width=0.8\columnwidth,clip]{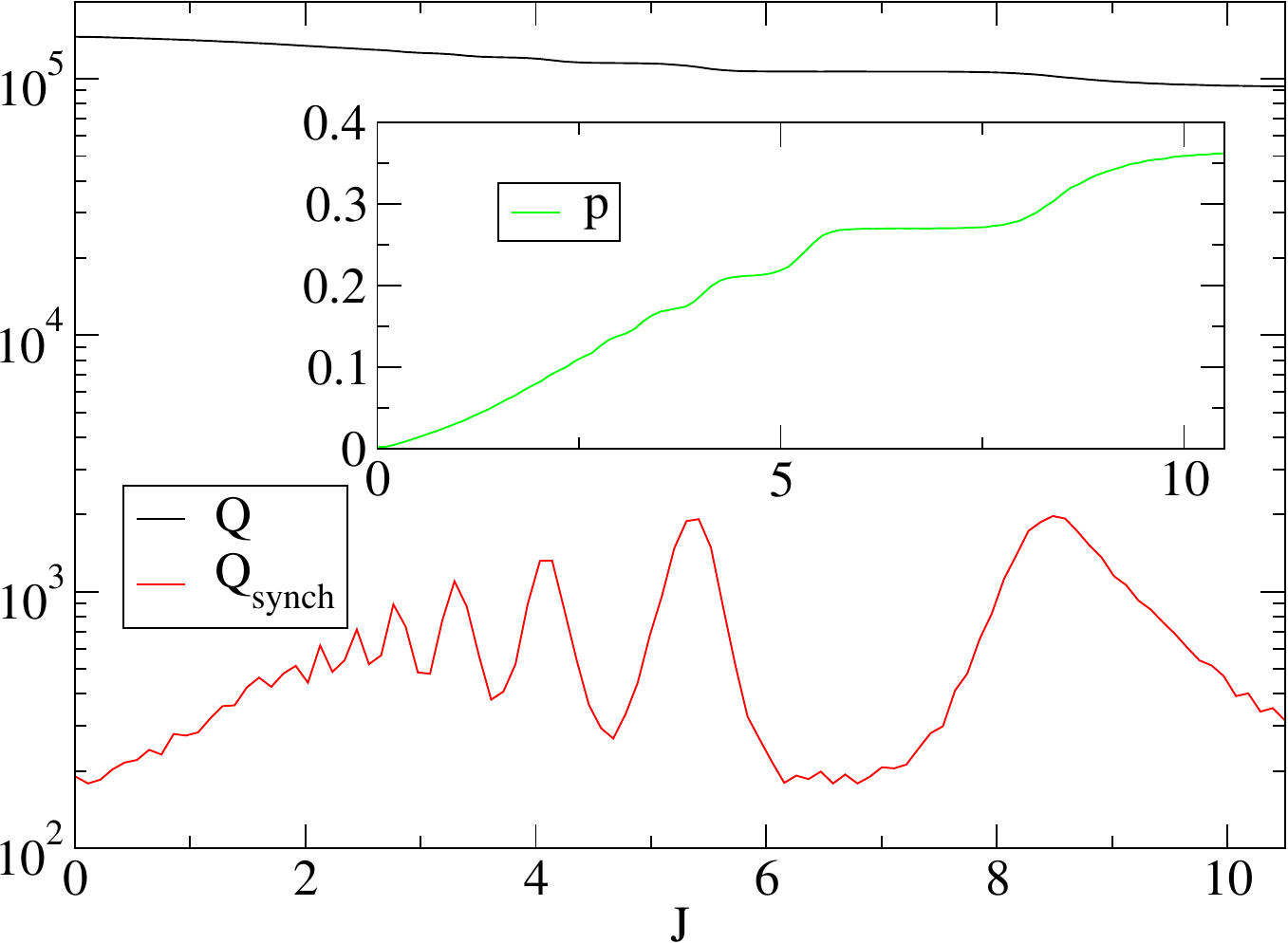}
    \caption{
      The number $Q$ of clusters, the number $Q_{\mathrm{synch}}$ of clusters having size larger than 1, and the participation ratio $p$ (Inset)
      as functions of $J$ obtained
      via the MPA with random initial conditions and random noise in the range $[-1,1]$ applied to the network of the WordNet network. 
    Here we used $t_{\mathrm{max}}= 10^3$.}
    \label{wordnet_noisy}
  \end{figure}

For this network, we focused exclusively on the noisy version of the algorithm. Despite being almost {30} times larger than the US power-grid network, convergence was achieved with a relatively small number of iterations, $t_{\mathrm{max}}= 10^3$.
This faster convergence can be attributed to the markedly different topology: the presence of highly connected hubs renders this network locally tree-like (at least effectively),
a condition under which message-passing algorithms are known to perform particularly efficiently.
The resulting behaviors of $Q$, $Q_{\mathrm{synch}}$ and $p$ (Inset) as functions of $J$
are reported In Fig. \ref{wordnet_noisy}.
As expected, also in this case, $Q$ and $p$ are monotone functions of $J$
exhibiting smooth plateaus. While the interpretation of the resulting partition in semantic terms is potentially very rich, it lies {beyond} the scope of the present work.

\section{Conclusions and Perspectives}
In this work, we addressed the problem of identifying stable clusters of synchronized nodes in large complex networks when the underlying microscopic dynamics is unknown or stochastic. By adopting a probabilistic perspective and introducing a surrogate dynamics based on the message-passing algorithm, we proposed an unbiased and scalable framework that avoids ad hoc assumptions on the system dynamics.

The method relies on a reduction of node states to binary variables and exploits the collective behavior emerging across critical points of an effective Ising-like model. This approach allows the identification of dynamically coherent clusters while scaling efficiently with network size, making it suitable for the analysis of large real-world networks. Our results highlight the key role of topologically equivalent nodes acting as nucleation centers for synchronization,
giving rise to rich collective scenarios. Notably, abrupt desynchronization transitions were observed
even in simple graphs, without the need to invoke higher-order interactions. When stochastic noise is included, however,
a more realistic synchronization scenario sets in.
In fact, when noise is present, the transitions across critical points get smoother, and the function $Q(J)$ and $p(J)$ become monotone and characterized by the emergence of plateaus,
albeit at the cost of larger coupling strengths. 

Finally, the application of the proposed method to the US power-grid and WordNet networks demonstrates its practical feasibility on large real-world systems with markedly different topologies, while confirming that the observed synchronization and desynchronization patterns primarily reflect structural properties of the underlying networks rather than specific physical dynamics.

A limitation of the present framework stems from the binary nature of the state variables (a coarse-grained description),
which tends to merge multiple distinct groups of topologically equivalent nodes into relatively large synchronized clusters (low resolution).
Extending the approach to multi-state variables represents a natural and promising direction for future work, as it would provide higher resolution,
while preserving the probabilistic and message-based character of the framework.

\section*{Data availability}
The Fortran codes used to generate the results presented in this work are publicly available and permanently archived on Zenodo at
\href{https://doi.org/10.5281/zenodo.18235744}{https://doi.org/10.5281/zenodo.18235744}.
{The archive contains three programs.
The first generates synthetic graphs with a desired number of groups of TE nodes, discards 
isolated groups, and returns a connected graph. The second runs the MPA on graphs generated by the first program
to detect synchronized clusters and compares them with the corresponding TE groups.
The third runs the MPA to detect synchronized clusters for an arbitrary input graph (without prior information on TE groups).}

\begin{acknowledgments}
This work was supported by the National Council for Scientific and Technological Development (CNPq), Brazil, under the Universal Grant (process no. 409180/2023-8).
\end{acknowledgments}

\appendix

\section{\label{AppA}Diagram of the algorithm}
\begin{widetext}

\begin{table}[h!]
  \begin{center}
    \caption{A synthetic description of our algorithm for partitioning a given graph $\mathcal{G}$ into $Q$ clusters of stable synchronized nodes. Note that
      $u_{i\to\j}$ is a directional message and does not make sense to call it incoming our outcoming. However, in order to save memory, we introduce
      the ``local matrices'' $u^{\mathrm{(IN)}}(l,i)=u_{j(l)\to i}$ and $u^{\mathrm{(OUT)}}(l,i)=u_{i\to j(l)}$
      where the former index runs in the set $\{1,\ldots,k(i)\}$ (according to an arbitrary but established order),
      $k(i)$ being the number of neighbors of the node $i\in V$ and $j(l)\in V$ stands for the $l$-th first neighbor of $i$. To save further memory,
      in our actual FORTRAN program 
      the above local matrices are replaced by suitable vectors (vectorizations). The value of $J_{\mathrm{min}}$ should be chosen as close as possible
      to the critical value $J_c=J_1$. For very large networks, in order to safe time, it is convenient to
      first run another version of the below algorithm in which line \ref{initialize} is replaced by
      ``Initialize $|E|$ incoming messages of each node with the value 0.1''. This version will converge much faster allowing a fine evaluation of $J_c$ to be used then
      in the original algorithm as $J_{\mathrm{min}}$.
    }
    \vspace{0.5cm}
    \label{table0}
    \begin{tabular}{|p{15cm}|}
      \hline
    \begin{enumerate}
    \item \quad Read $\mathcal{G}=(V,E)$ via file named ``edges.txt'' 
    \item \quad For $J\in\{J_{\mathrm{min}},\ldots,J_{\mathrm{max}}\}$ 
    \item\label{initialize} \quad \quad Initialize $N$ random fields from uniform distributions in $[-h^*,h^*]$
    \item\quad \quad Initialize $E$ incoming messages from uniform distribution in $[-1,1]$
    \item \quad \quad For $t\in\{0,\ldots,t_{\mathrm{max}}\}$ 
    \item \quad \quad \quad Compute $|E|$ outgoing messages via Eq. (\ref{BP01}) 
    \item  \quad \quad \quad Read the new $|E|$ incoming messages from the $|E|$ outgoing messages
    \item \quad \quad \quad Check stationarity of all incoming and outgoing messages via Eq. (\ref{disttime})
    \item \quad \quad If stationarity holds for all messages GOTO \ref{statline}     
    \item \quad \quad End-For 
    \item \quad \quad Return ``non stationarity'' 
    \item\label{statline} \quad \quad Continue
    \item \quad \quad Compute local magnetization of each node via Eq. (\ref{mag}) 
    \item \quad \quad Order values of local magnetizations 
    \item \quad \quad Find $Q$ clusters via Eqs. (\ref{dist}) 
    \item \quad \quad Compute metrics such as mean and variance of $N^{(l)}$, $k^{(l)}$, $k^{(\mathrm{IN};l)}$, $k^{(\mathrm{OUT};l)}$ 
    \item \quad \quad Return full cluster decomposition and its metrics
    \item\label{statline2} \quad \quad Continue
    \item \quad End-For      
    \end{enumerate}\\
    \hline
    \end{tabular}
  \end{center}
\end{table}

\end{widetext}


\end{document}